\documentclass{eas}
\usepackage{graphicx}
\begin{document}

\title{Simulation of \\ the Directional Dark Matter Detector (D$^3$) \\ and \\
Directional Neutron Observer (D{\small i}NO) }
\author{I. Jaegle$^1$, H. Feng$^1$, S. Ross$^1$, J. Yamaoka$^1$, S.E. Vahsen}
\address{Department of Physics and Astronomy, University of Hawaii, 2505 Correa Rd., Honolulu, HI 96822\\
contact person: Igal Jaegle - igjaegle@gmail.com.}
%
%
\begin{abstract}

Preliminary simulation and optimization studies of the Directional Dark Matter Detector and the Directional Neutron Observer are presented. These studies show that the neutron interaction with the gas-target in these detectors is treated correctly by GEANT4 and that by lowering the pressure, the sensitivity to low-mass WIMP candidates is increased. The use of negative ion drift might allow us to search the WIMP mass region suggested by the results of the non-directional experiments DAMA/LIBRA, CoGeNT and CRESST-II.
\end{abstract}
\maketitle
\section{Introduction}

Monte Carlo simulations are not only essential tools for the comparison of theory
and experiment in physics, but also for the design and optimization of detectors.
In these proceedings, we present preliminary simulation studies
of Time Projection Chambers (TPCs), where the drift charge is amplified with
 Gas Electron Multipliers (GEMs) and detected with pixel electronics. This TPC configuration is described
 in these proceedings of S. Vahsen [\cite{Vahsen_2011}] and also in [\cite{Vahsen_2008}]. These technologies should allow improved gas-target detectors, where the ionization in the target gas is detected with low noise, good position
and time resolution, and high efficiency. These features allow us to measure the momentum and energy of charged particles and indirectly of fast neutrons by measuring charged recoils procured when they scatter elastically off the nuclei of the gas-target. TPCs with GEMs and pixels may also allow
dark matter searches with improved sensitivity and background rejection that exploit the predicted twelve-hour oscillation Weakly Interacting Massive Particles (WIMPs), which results from the Earth's rotation [\cite{Lewin_1996},\cite{Directional}].

We will present ongoing simulation studies, including first sensitivity-estimates for a
Directional Dark Matter Detector (D$^3$) and a Directional Neutron Observer (DiNO)
based on these technologies. These proceedings are divided into four parts: the {\bf simulation strategy} is introduced, the {\bf preliminary simulation validation} of GEANT4 [\cite{GEANT4}] and of the WIMP cross section limit code are discussed, the {\bf design optimization} by varying the pressure and by using either the electron drift (ED) or the negative ion drift (NID) is presented, and finally the
{\bf preliminary results} are shown.

\section{Simulation strategy}

The D$^3$/DiNO Monte Carlo simulation toolkit is currently being assembled and will mostly rely on already available simulation programs: GEANT4 [\cite{GEANT4}], SRIM  [\cite{SRIM}] and GARFIELD [\cite{GARFIELD}]). The flow chart in Figure 1 shows the simulation steps
expected, which will perform the following tasks:

\begin{figure}
\begin{center}
  \includegraphics[height=.3\textheight]{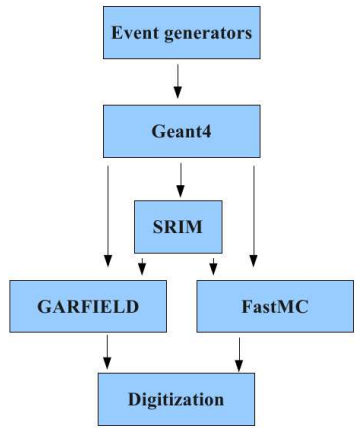}
  \caption{Flow chart showing the simulation steps expected for photons, WIMPs, nucleons, nuclei, leptons and mesons.}
\end{center}
\end{figure}

\begin{itemize}
\item{An event generator will simulate the signal and background sources. E.g., for the WIMP case: the predicted WIMP velocity distribution
    for the signal and cosmic-ray-induced radiation and radiation emitted by detector material for the background.}
\item{An accurate geometry and materials description.}
\item{The interaction between the incoming particle and the target gas or/and detector materials will be modeled using either the GEANT4 [\cite{GEANT4}]  physic classes, SRIM [\cite{SRIM}],  or GARFIELD  [\cite{GARFIELD}], depending the particle type, which may lead to the creation of ionization along the trajectory of the incoming particle or/and along the recoil product of the incoming particle scattering off gas-nuclei. SRIM[\cite{SRIM}] and GARFIELD[\cite{GARFIELD}] can model theses processes fairly well.}
\item{Then the electrons [\cite{ED}] (or negative ions [\cite{NID}]) drift under the influence of the electric field. Negative ions can be formed in the case the electrons from the ionization can attach themselves to the gas-molecule to form negative ions). GARFIELD or a fast Monte Carlo simulation using the gas properties calculated by MAGBOLTZ [\cite{MAGBOLTZ}] can simulate the drift of ionization (electrons, or negative ions in the case of negative ion drift), towards the GEMs with a constant velocity in a homogeneous electric field. The GEMs which then amplify the signal. In an area of high field near the GEMs, the electrons detach from the negative ion. Therefore a normal avalanche occurs both in the case of electrons and negative ions. We plan to model the GEMs with a parameterized simulation.}
\item{The digitization software then simulates how the resulting avalanche-charge is detected by the electronic readout, in our case pixel electronics [\cite{Vahsen_2008}] .}
\end{itemize}

\begin{figure}
\begin{center}
  \includegraphics[height=.3\textheight]{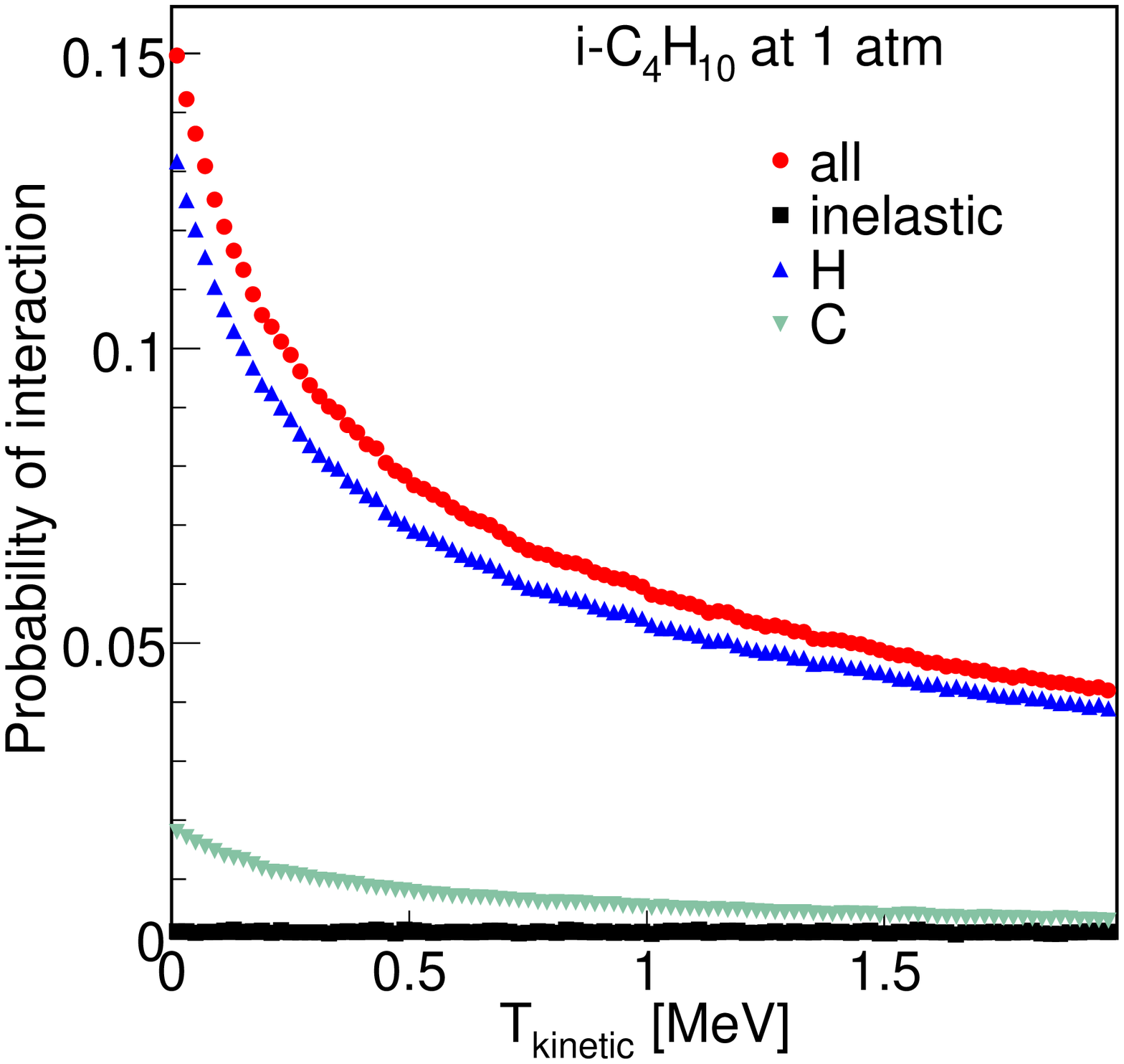}
  \includegraphics[height=.3\textheight]{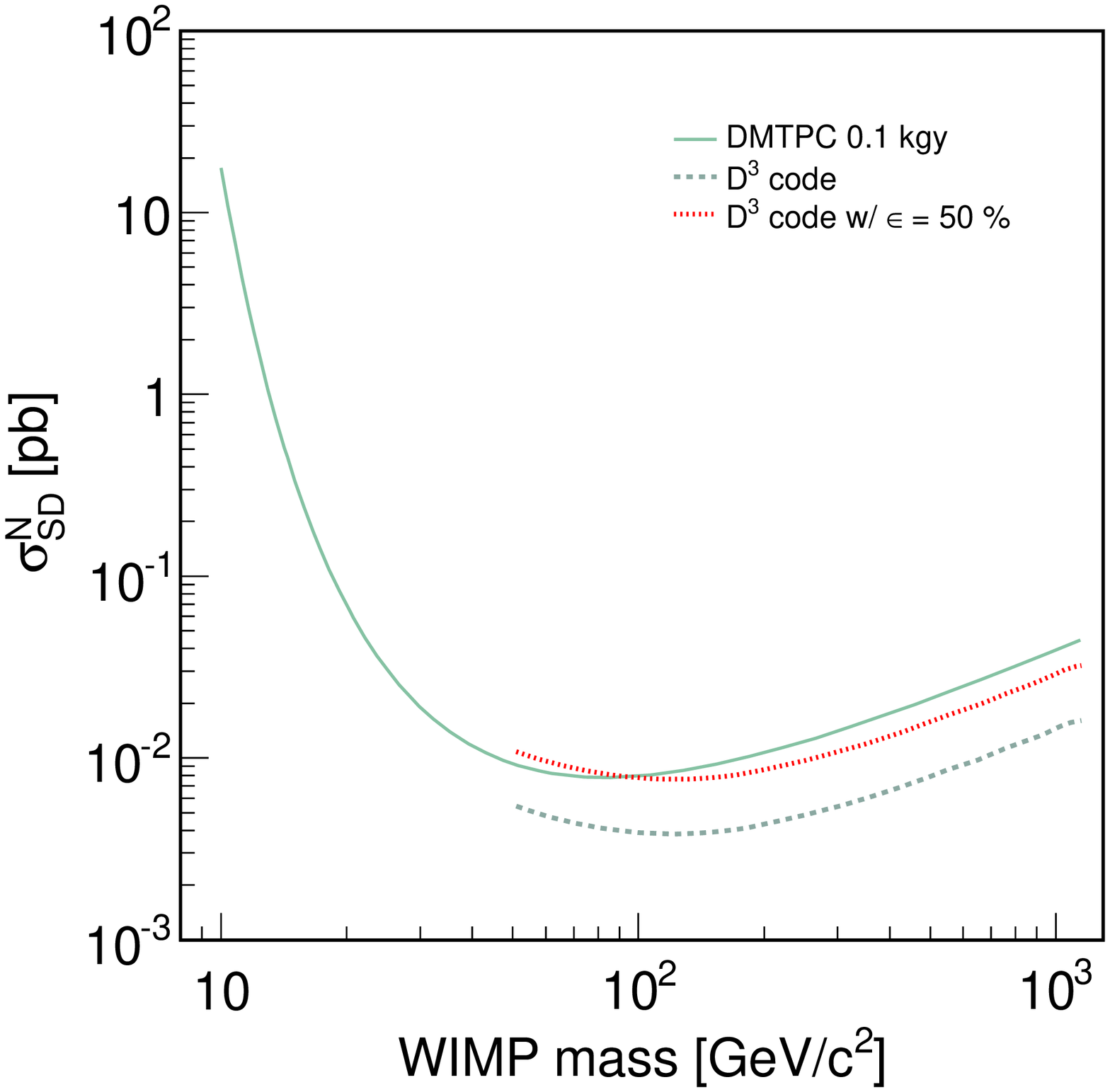}
  \caption{Left:  Probability of interaction, as a function of the neutron kinetic energy, between the neutron and iso-butane gas-nuclei
(red point and black square by isolating the inelastic scattering contribution) and between the neutron and the individual atomic nuclei
 (blue triangle up hydrogen, green triangle down carbon). Right: Spin dependent cross section limit as function of the WIMP
mass for DMTPC [\cite{DMTPC}] (green line) and as estimated by our fast simulation, assuming 100 $\%$ (dashed green line) and 50 $\%$ (red line) light collection
efficiency.}
\end{center}
\end{figure}

\section{Simulation validation}

\subsection{Neutron}

The probability of interaction between a neutron and the gas-target (50 cm length of iso-butane at 1 atm) has been simulated as function of the neutron kinetic energy (see Figure 1 left) with the
transport code  GEANT4 [\cite{GEANT4}]. This probability has also been calculated for 1-MeV neutrons by using the Low Energy Nuclear Data (LEND)
 [\cite{LEND}]. There is 0.11 $\%$ per centimeter probability  at 1 atm of 1-MeV neutron interacting with a Hydrogen atom belonging to the iso-butane-molecule. This analytical calculation is in good agreement with the GEANT4 [\cite{GEANT4}] calculation, which also uses the LEND [\cite{LEND}] for the cross section values, and gives 0.114 $\%$ per centimeter.

\subsection{WIMP}

A fast Monte Carlo simulation for estimating WIMP cross section limits has been implemented following the instructions of Lewin et al. [\cite{Lewin_1996}], using SRIM [\cite{SRIM}] for the track length simulation and MAGBOLTZ [\cite{MAGBOLTZ}] for the target-gas properties. The results have been compared to the DMTPC [\cite{DMTPC}] published limit by applying the DMTPC setup parameters. A fair agreement is found, as illustrated by Figure 2 (right).

\section{Design optimization}

The design optimization consists of finding the optimum pressure for a directional dark matter detector, D$^3$, when there is a good trade off between the target mass and track length so that the directional sensitivity is maximized.
The volume is kept fixed by considering a detector with a one square meter readout plane and a maximum drift length of 33.33 cm. 

Several conditions are imposed on the track to ensure that the projection of the track, with a length L, on the two-dimensionally segmented pixel chip
readout plane can be exploited to extract the directionality:
\begin{itemize}
\item{L $>$ 3 $\sigma_{xy}$ where $\sigma_{xy}$ is the transverse diffusion}
\item{L $>$ 3 $\times$ GEM holes spacing}
\item{energy threshold on the primary ionization energy of 1-keV that corresponds approximately to 40 electrons detected}
\end{itemize}

An approximate quenching factor, estimated with SRIM, was also included.

\begin{figure}
\begin{center}
   \includegraphics[height=.3\textheight]{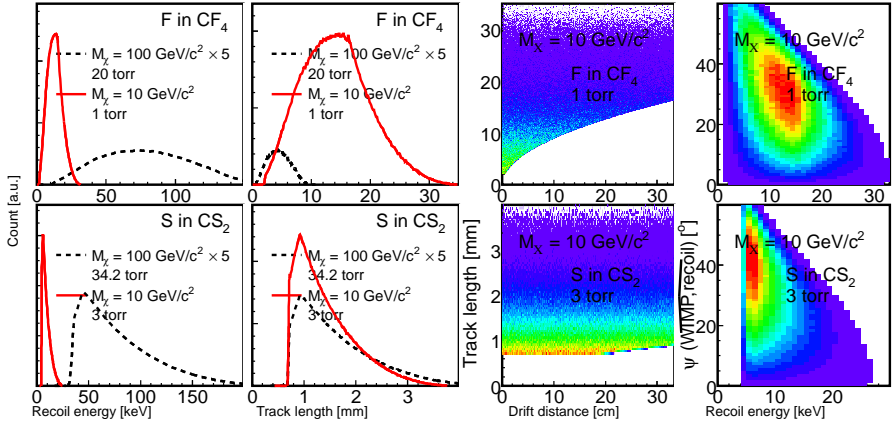}
  \caption{Fluorine in CF$_4$(- top row) and sulfurine in CS$_2$(-bottom row): recoil (kinetic) energy distribution (first column), recoil track length distribution (second column), track length versus drift distance for 10 GeV/c$^2$ WIMP mass (third column), angle between the incoming WIMP and the scatter recoil versus recoil (kinetic) energy for 10 GeV/c$^2$ WIMP mass (fourth column). The directionality conditions are applied. For the recoil energy and track length distributions, the distributions for 100 GeV/c$^2$ WIMP mass are scaled by a factor 5.}
\end{center}
\end{figure}

ED and NID, when the ratio between the electric field and pressure is kept constant, the transverse diffusion scales with the square root of the pressure (or the number density). Similarly, the Townsend and
 attachment coefficients scale with the pressure. Therefore if we decrease the pressure the transverse diffusion will get worse and the
probability of being in a multiplication regime increases as well. However, the track length scales with the pressure. If we decrease the
pressure the track length increases.

The figure of merit (FOM) is given by the number of WIMP recoils expected to result in a reconstructible track. It has been calculated for several gases
(H$_2$, C$_2$H$_6$, C$_4$H$_{10}$, $^4$He, CF$_4$, $^{40}$Ar and $^{132}$Xe for the ED case and  CS$_2$ for the NID case) and for two WIMP masses (10 GeV/c$^2$ and 100  GeV/c$^2$). The figure of merit can be written as:

\begin{eqnarray}
\frac{dFOM}{dT_R}(P) = \frac{\mu_A^2}{\mu_N^2} . \rho(P) . V \frac{d}{dT_R} \Gamma^A . \frac{ \int_0^{z_{max} = 33.33 cm} L_{L>L_0(P)}(T_R,P) dz}{L(T_R,P)}
\end{eqnarray}
where:
\begin{itemize}
\item{$\mu_A$ and $\mu_N$ are the nucleus and the nucleon reduced mass, respectively}
\item{$\rho$ is the number of gas-molecules per cubic centimeter}
\item{V is the target gas volume}
\item{$\Gamma^A = F^2(qr_n)I$ with $F^2(qr_n)$ the form factor as defined in [\cite{Lewin_1996}] and I=A$^2$ for the spin independent case (SI) or I $\propto$ J(J+1) for the spin dependent case (SD)}
\item{L is the track length and L$_0$ is the track length ``threshold'' derived by the three conditions described above. L$_0$ is not a constant and is changing for each pressure}
\item{P is the pressure}
\item{T$_R$ is the kinetic energy of the recoil nucleus}
\end{itemize}

\begin{figure}
\begin{center}
  \includegraphics[height=.3\textheight]{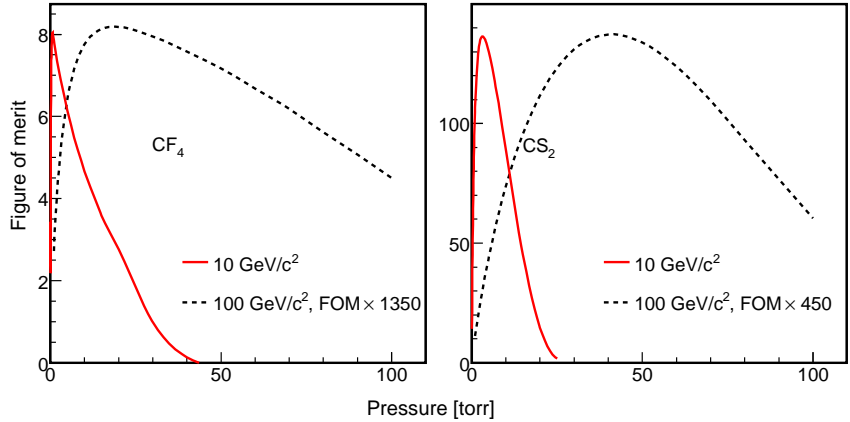}
  \caption{Figure of merit versus pressure for CF$_4$ (left) and CS$_2$ (right) calculated respectively for 10 GeV/c$^2$ (red, full line) and 100 GeV/c$^2$ WIMPs (black, dashed line which is scaled by factor 1350 for CF$_4$ and 450 for CS$_2$).}
\end{center}
\end{figure}

Example distributions (track length and recoil energy) corresponding to the optimum pressures found for WIMP masses of 10 GeV/c$^2$ and
 100 GeV/c$^2$, after the directionality conditions are applied, can be seen in Figure 3. Figure 3 also shows the track length versus the drift distance and the angle between the
 incoming WIMP and the recoil nucleus versus the recoil energy. For the 10 GeV/c$^2$ WIMP mass case, the most probable 
track length and recoil energy are: 15 mm, 13 keV and 0.8 mm, 5.8 keV for respectively fluorine in CF$_4$ and sulfurine in CS$_2$. As expected in the case of NID, the sulfurine recoil in CS$_2$ have fairly short track at the optimum pressure. The cutoff at 0.7 mm in the track length distribution for CS$_2$ corresponds to the GEM holes spacing requirement. 

Finally, Figure 4 shows the resulting figure of merit for fluorine recoils in CF$_4$ and sulfurine recoils in CS$_2$. The optimum pressures to detect a 10 GeV/C$^2$ WIMP mass are 1 torr and 3 torr, respectively for CF$_4$ and
CS$_2$. It appears that only CS$_2$ among all the gases tested is competitive and can run at such low pressure. For example, MAGBOLTZ [\cite{MAGBOLTZ}] indicates that at 1 torr CF$_4$, one will be in a multiplication regime for an electric field of few 10's V/cm, above 100's V/cm at 1 torr one is far away the thermal limit. The
gases with light nuclei can also run for the low pressure given by the FOM but are not competitive. For a 100 GeV/c$^2$ WIMP mass only CF$_4$ and CS$_2$ are competitive.

For CS$_2$-NID, a very low transverse diffusion of 50 $\mu m / \sqrt{cm}$ at 80 torr and an electric field of 1 kV/cm have been reported in [\cite{NID}] and is expected to run also at very low pressure with an electric field of few 10's V/cm according to [\cite{NID}]. While for
comparison CF$_4$ has a transverse diffusion of 150  $\mu m / \sqrt{cm}$ at 80 torr and an electric field of 2 kV/cm.


\section{Preliminary reach plots}

\begin{figure}
\begin{center}
  \includegraphics[height=.3\textheight]{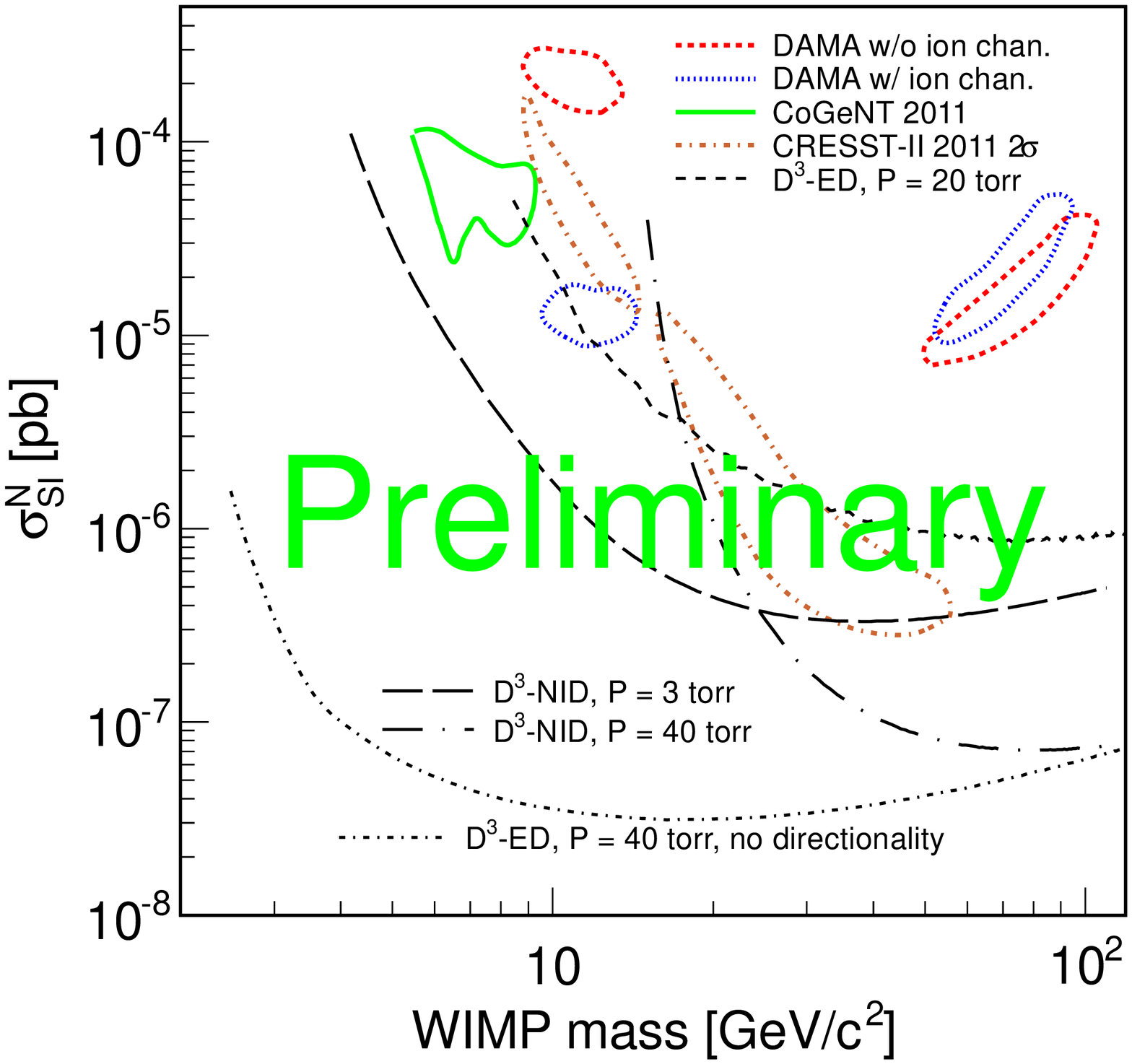}
  \includegraphics[height=.3\textheight]{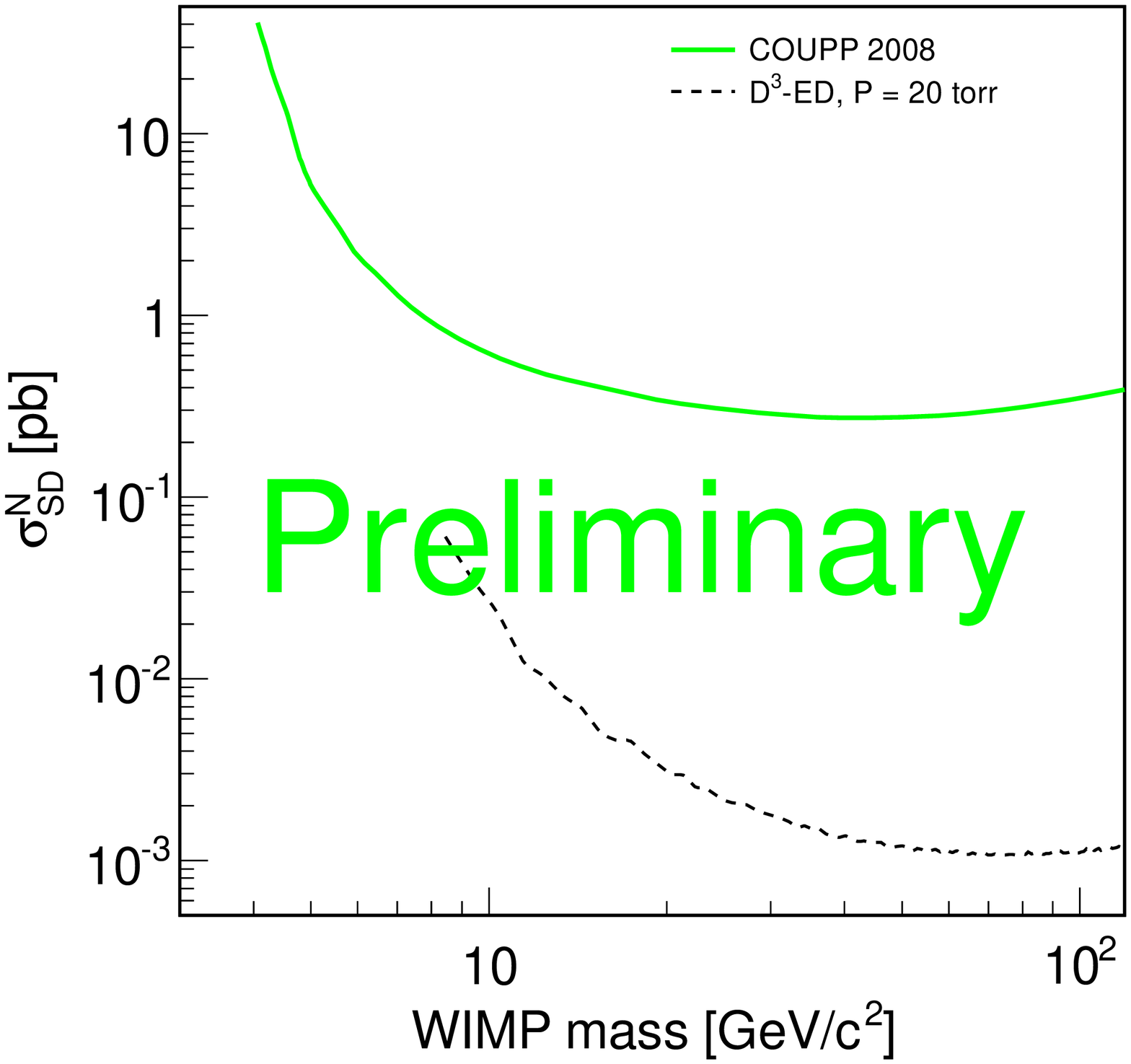}
  \caption{D$^3$ cross section limit as a function of the WIMP mass for one recoil produced by a WIMP detected in three m$^3$. The detector is divided into nine sub-detectors with a
maximum drift distance of 33.33 cm for ED-CF$_4$ and NID-CS, the SI case on the left and for the SD case on the right. The D$^3$ reach plot is compared to the non-directional experiments DAMA/LIBRA [\cite{DAMA}], CoGeNT [\cite{CoGeNT}] and CRESST-II [\cite{CRESST}] for the SI case and to COUPP[\cite{COUPP}] for the SD case.}
\end{center}
\end{figure}

Figure 5 shows preliminary reach plots for WIMP search with three m$^3$ of CS$_2$ or CF$_4$, for three years of exposure. The detector is divided into nine sub-detectors with a maximum drift length of 33.33 cm. The WIMP velocity distribution used is Maxwellian, the dark matter density is 0.3 GeV/c$^2$/cm$^3$, and the escape velocity is
530 km/s.

 The sensitivity to low mass WIMPs is increased for low pressure as illustrated by the CS$_2$ results which can be seen in Figure 5 (left). If the pressure is
increased at the cost of directionality, the sensitivity is drastically improved as expected (see D$^3$-ED, P = 40 torr no directionality curve in Figure 5 - left) so that after only a month of exposure the reach will be already below he region suggested by DAMA/LIBRA [\cite{DAMA}] and CoGeNT [\cite{CoGeNT}].
In principle a carefully designed TPC could decrease the pressure or even change the gas to CS$_2$, once a possible WIMP candidate is detected. Then by switching to directional mode, after three years DAMA/LIBRA [\cite{DAMA}], CoGeNT [\cite{CoGeNT}] and CRESST-II [\cite{CRESST}] can be either excluded or confirmed unambiguously.

\section{Conclusion}

Preliminary simulation and optimization results of the Directional Dark Matter Detector and the Directional Neutron Observer have been presented. These results show that the neutron interaction with the gas is treated correctly by GEANT4 [\cite{GEANT4}] and that by lowering the pressure, the sensitivity to low-mass WIMP candidates is
increased. The use of negative ion drift might allow us to search the WIMP mass region suggested by results of the non-directional experiments DAMA/LIBRA [\cite{DAMA}], CoGeNT [\cite{CoGeNT}] and CRESST-II [\cite{CRESST}].



\begin{thebibliography}{99}


\bibitem[1]{Vahsen_2011}[1] S.E. Vahsen, The Directional Dark Matter Detector (D$^3$) CYGNUS 2011 proceedings.
\bibitem[2]{Vahsen_2008}[2] T. Kim et al, NIM A 589 (2008) 173~@~S184.
\bibitem[3]{Lewin_1996}[3] J.D. Lewin, F.P. Smith, Astroparticle Physics 6 (1996) 87-112.
\bibitem[4]{Directional}[4] S. Ahlen, N. Afshordi, J. Battat, J. Billard, N. Bozorgnia, et al., The case for a directional dark matter detector and the status of current experimental efforts, Int.J.Mod.Phys. A25 (2010) 1~@~S51. arXiv:0911.0323, doi:10.1142/S0217751X10048172.
\bibitem[5]{GEANT4}[5] "GEANT4" geant4.cern.ch.
\bibitem[6]{SRIM}[6] "SRIM" M. D. Z. James F. Ziegler, Jochen P. Biersack, SRIM: The Stopping and Range of Ions in Matter, Lulu Press Co., 2009. srim.org.
\bibitem[7]{GARFIELD}[7] "GARFIELD" R. Veenhof, GARFIELD, recent developments, Nucl.Instrum.Meth. A419 (1998) 726~@~S730. doi:10.1016/S0168-9002(98)).
008511. http://garfield.web.cern.ch/garfield/.
\bibitem[8]{ED}[8] F. Sauli, Drift and diffusion of electrons in gases: a compilation (with an introduction to the use of computing programs) CERN, 1984. - 127 p.
\bibitem[9]{NID}[9] C. J. Martoff, Negative Ion TPC for WIMP AstronomySNIC Symposium, Stanford, California -- 3-6 April 2006.
\bibitem[10]{MAGBOLTZ}[10] http://consult.cern.ch/writeup/magboltz/.
\bibitem[11]{LEND}[11] National Nuclear Data Center, http://www.nndc.bnl.gov/exfor/endf00.jsp.
\bibitem[12]{DMTPC}[12] "DMTPC" The DMTPC Detector. G. Sciolla et al., e-Print arXiv:0811.2922.
\bibitem[13]{DAMA}[13] "DAMA/LIBRA" R. Bernabei, P. Belli, F. Cappella, R. Cerulli, C. Dai, et al., New results from DAMA/LIBRA, Eur.Phys.J. C67 (2010) 39~@~S49. arXiv:1002.1028, doi:10.1140/epjc/s10052-010-1303-9.
\bibitem[14]{CoGeNT}[14] "CoGenT" C. Aalseth, P. Barbeau, J. Colaresi, J. Collar, J. Leon, et al., Search for an Annual Modulation in a P-type Point Contact Germanium Dark Matter Detector, Phys.Rev.Lett. 107 (2011) 141301. arXiv:1106.0650.
\bibitem[15]{CRESST}[15] "CRESST-II" G. Angloher, M. Bauer, I. Bavykina, A. Bento, C. Bucci, et al., Results from 730 kg days of the CRESST-II Dark Matter Search, arXiv:1109.0702.
\bibitem[16]{COUPP}[16] "COUPP" Behnke et al., Science 319 p. 933 (2008).
\end{thebibliography}
\end{document}